\documentclass[lettersize,journal]{IEEEtran}
\usepackage{amsmath, amssymb, amsfonts, mathtools} 
\usepackage{amssymb, bm}
\usepackage{algorithmic}
\usepackage{algorithm}
\usepackage{array}
\usepackage[caption=false,font=normalsize,labelfont=sf,textfont=sf]{subfig}
\usepackage{textcomp}
\usepackage{stfloats}
\usepackage{url}
\usepackage{verbatim}
\usepackage{graphicx}
\usepackage{cite}
\usepackage{subcaption}
\usepackage{xcolor}
\usepackage{booktabs}
\usepackage[table]{xcolor}
\usepackage{textcase}
\newcommand{\grey}{\cellcolor{blue!9}}

\def\a {\mathbf{a}}
\def\b {\mathbf{b}}

\def\d {\mathbf{d}}

\def\p {\mathbf{p}}

\def\u {\mathbf{u}}

\def\x {\mathbf{x}}
\def\y {\mathbf{y}}
\def\z {\mathbf{z}}

\def\A {\mathbf{A}}
\def\B {\mathbf{B}}
\def\C {\mathbf{C}}
\def\D {\mathbf{D}}
\def\E {\mathbf{E}}
\def\F {\mathbf{F}}
\def\G {\mathbf{G}}
\def\H {\mathbf{H}}
\def\I {\mathbf{I}}

\def\K {\mathbf{K}}
\def\L {\mathbf{L}}
\def\M {\mathbf{M}}
\def\N {\mathbf{N}}
\def\O {\mathbf{O}}

\def\R {\mathbf{R}}

\def\T {\mathbf{T}}
\def\U {\mathbf{U}}
\def\V {\mathbf{V}}
\def\W {\mathbf{W}}
\def\X {\mathbf{X}}
\def\Y {\mathbf{Y}}
\def\Z {\mathbf{Z}}

\def\hA {\hat{\mathbf{A}}}
\def\hB {\hat{\mathbf{B}}}

\def\hd {\hat{\mathbf{d}}}
\def\hD {\hat{\mathbf{D}}}

\def\hR {\hat{\mathbf{R}}}

\def\hI {\hat{\mathbf{I}}}

\def\bx {\bar{\mathbf{x}}}
\def\by {\bar{\mathbf{y}}}
\def\ba {\bar{\mathbf{a}}}

\def\bh {\bar{\mathbf{h}}}
\def\bf {\bar{\mathbf{f}}}

\def\bA {\bar{\mathbf{A}}}
\def\bB {\bar{\mathbf{B}}}
\def\bY {\bar{\mathbf{Y}}}

\def\tT {\tilde{\mathbf{T}}}
\def\tI {\tilde{\mathbf{I}}}

\def\rT {\mathrm{T}}
\def\rH {\mathrm{H}}
\def\rF {\mathrm{F}}

\def\rr {\mathrm{r}}

\newcommand{\sint}[1]{\int_\mathbf{u} #1 \mathrm{d}\mathbf{u}}
\newcommand{\sintt}[1]{\int_{\mathbf{u}'} #1 \mathrm{d}\mathbf{u}'}
\newcommand{\ssint}[1]{\int_\mathbf{u}\int_{\mathbf{u}'} #1 \mathrm{d}\mathbf{u}'\mathrm{d}\mathbf{u}}

\newcommand{\Tr}[1]{\mathrm{Tr}\left[ #1 \right]}
\newcommand{\Ex}[1] {\mathbb{E}\left[ #1 \right]}

\DeclareMathOperator*{\argmin}{arg\,min}

\begin{document}

\title{Generalised Transcoding Framework for Arbitrary Spatial Audio Capture and Playback Formats}

\author{Archontis Politis, Janani Fernandez, and Leo McCormack
\thanks{Archontis Politis and Janani Fernandez are with the Faculty of Information Technology and Communication Sciences, Tampere University, Finland. \\ \indent Leo McCormack is with the Department of Information and Communications Engineering, Aalto University, Espoo, Finland.}
}

\markboth{Preprint. Under review for possible publication in an IEEE journal.}%
{Shell \MakeLowercase{\textit{et al.}}: A Sample Article Using IEEEtran.cls for IEEE Journals}


\maketitle

\begin{abstract}
This article introduces a unified framework for the parametric analysis and reproduction of spatial sound scenes captured either as Ambisonic signals or as raw microphone array signals. The proposed method estimates time-frequency-dependent spatial metadata that characterises a variable number of primary source components and an ambience component with its own angular power distribution, whose parameters fit the observed spatial covariances of the captured signals. This metadata is used to construct spatial covariances of the target playback formats, which are then used to derive optimal mixing matrices for transcoding the scene for playback over the target reproduction system. The method additionally handles independent rotations of both capture and playback setups. Real-time implementations of the method and other existing state-of-the-art parametric renderers are compared in a listening test using simulated scenes from Ambisonic, spherical, and head-worn arrays. The results highlight perceptual benefits of the proposed framework across a diverse range of content and receiver configurations, particularly for lower-order and geometrically constrained microphone arrays.
\end{abstract}


\section{Introduction} 

\IEEEPARstart{F}{lexible} and spatially accurate capture and reproduction frameworks are becoming increasingly sought after within augmented, virtual, mixed, and extended reality (AR/VR/MR/XR), immersive media, digital‑twin, and telepresence applications. Historically, large‑scale spatial audio reproduction has relied on fixed multichannel loudspeaker setups (e.g., 5.1 home cinema systems), delivering handcrafted object‑based content or real‑world recordings that did not necessarily exhibit the same spatial characteristics as the original scene. Today, binaural rendering over headphones and earbuds dominates consumer playback, and high reproduction accuracy is essential for many emerging applications. Modern systems also incorporate head‑tracking to improve immersion and externalisation \cite{best2020sound}, requiring dynamic time‑variant rendering. Smart immersive loudspeakers and multi‑driver soundbars are likewise gaining popularity, offering spatial audio within domestic environments with flexible hardware placement. Therefore, modern spatial rendering systems must accommodate a wide range of playback setups and use cases.

To capture spatial sound scenes, professional engineers have traditionally used custom‑designed microphone arrays optimised for specific playback formats \cite{lee2021multichannel}, or spherical microphone arrays (SMAs) that provide uniform spatial capture resolution and can be subsequently converted to target formats using mixing matrices (or a matrix of filters) \cite{backman2003microphone, ono2013portable, chen1992external, li2006headphone}. SMAs are naturally the more flexible option, particularly since they are also well-suited for transforming captured sound scenes into Ambisonics \cite{gerzon1973periphony}; a playback‑agnostic spatial audio format widely adopted in major media platforms (YouTube, Facebook 360) and recent standards (MPEG‑H, MPEG‑I, Eclipsa, IVAS). Linear decoders are then often used to reproduce Ambisonic scenes for binaural \cite{schorkhuber2018binaural} or loudspeaker playback \cite{zotter2012all}.

The most common professional SMA for Ambisonics capture uses four microphones arranged in a tetrahedral fashion, enabling first‑order Ambisonics (FOA) linear encoding up to the spatial aliasing frequency \cite{rafaely2015fundamentals}. FOA also remains the most widely used resolution within media platforms. Perceptual studies have shown, however, that linearly decoding FOA introduces coherent spreading of directional sounds (causing localisation blur) and ambience (leading to comb‑filtering and reduced envelopment) \cite{bertet2013investigation, avni2013spatial, fernandez2024investigating}. Higher‑order Ambisonics (HOA) SMAs mitigate these issues, but often rely on lower‑quality capsules \cite{bates2016comparing}, or become prohibitively expensive, while in any case only offering higher‑order components over limited frequency bandwidths \cite{moreau20063d}.

Signal‑dependent rendering strategies offer an alternative path. Parametric methods \cite{pulkki2017first, berge2010high, politis2018compass, mccormack2019parametric, schorkhuber2019linearly} adopt sound field models, estimate time-frequency‑dependent spatial parameters, and use them to synthesise perceptually enhanced playback signals. Directional energy is typically collapsed toward analysed directions to reduce localisation blur, while ambient components are decorrelated to preserve envelopment and avoid comb‑filtering. Parametric frameworks also enable intuitive sound‑field manipulations; such as listener translation, directional warping, and directional loudness controls \cite{mccormack2021parametric, fernandez2022spatial}. Sparse‑recovery approaches \cite{wabnitz2011upscaling, birnie2021mixed} can sharpen directional content without explicit sound‑field models, but generally neglect diffuse ambience and require computationally expensive iterative solvers. Recent deep neural network (DNN) based end‑to‑end Ambisonic upscaling and reproduction solutions \cite{routray2019deep, wang2022up, nawfal2024ambisonics} show promise, but perceptual evaluations remain limited, comparisons with other established signal-dependent methods are missing, and such systems typically lack interpretable run‑time controls or straightforward integration of personalised head-related transfer functions (HRTFs).

As a stark contrast to professional SMAs, modern consumer devices, such as smartphones, 360 degree cameras, hearables, and head‑mounted displays, typically contain only a few microphones placed suboptimally for linear Ambisonic encoding. As a result, device‑captured FOA scenes often suffer from spatial aliasing across much of the audible frequency range \cite{mccormack2022parametric}; therefore, even an optimal Ambisonics decoding solution would not be able to deliver a faithful rendition of the original captured sound scene. Two optimisation paths have therefore emerged: (i) improving Ambisonic encoding using parametric \cite{mccormack2022parametric}, sparse‑recovery \cite{bastine2022ambisonics}, or DNN‑based \cite{heikkinen2024neural} methods; or (ii) bypassing a potentially problematic Ambisonic encoding entirely, by instead mapping microphone signals directly to the target playback format.

Many studies have focused on optimising the direct microphone array to binaural playback task in particular. Linear approaches typically use least‑squares \cite{li2006headphone} or magnitude‑least‑squares \cite{deppisch2021end} optimisations to fit the directional capture characteristics of the device (described via array transfer functions), onto the target binaural directivity patterns (described by HRTFs). Additional constraints have also been explored specifically for head-worn array captures \cite{madmoni2020beamforming,berebi2024feasibility}. However, linear methods remain inherently limited by the number of microphones, with small numbers leading to perceptual issues similar to those observed when linearly decoding FOA. Therefore, parametric approaches for direct array‑to‑binaural rendering \cite{fernandez2022enhancing, mccormack2023six, stahl2024perceptual, berger2026performance} have also been explored, which have been shown to achieve improved perceived spatial accuracy over their linear counterparts.

In summary, signal‑dependent parametric methods have demonstrated high perceptual accuracy for Ambisonic encoding \cite{mccormack2022parametric}, Ambisonic decoding \cite{pulkki2017first, berge2010high, politis2017enhancement}, and direct microphone‑array‑to‑playback conversion \cite{fernandez2022enhancing, stahl2024perceptual}. They also support intuitive spatial manipulations \cite{mccormack2021parametric}, personalised binaural rendering by using arbitrary HRTFs, and real‑time operation within practical systems \cite{mccormack2019sparta}. However, no existing framework unifies Ambisonics‑based and raw microphone‑array‑based processing. Therefore, this article proposes Ge(n)eralised Coding and Multidirectional Parameterisation of Acoustic Sound Scenes (nCOMPASS), a generalised parametric framework that accepts either Ambisonics or raw microphone signals as input. nCOMPASS follows the same general philosophy as COMPASS \cite{politis2018compass} in aiming for flexible parameterisation and rendering across arbitrary playback configurations, while assuming a time-frequency sound‑field model of variable spatial complexity. However, unlike COMPASS, it avoids the explicit source and ambience spatial‑filtering operations that can lead to unstable behaviour in challenging conditions, such as closely spaced sources. Instead, nCOMPASS employs a more comprehensive formulation of the sound field directly in the spatial covariance domain, which is inherently more robust under such scenarios. By operating in this domain, the framework can forgo intermediate spatial filters and instead derive transcoding matrices that directly map the capture format to the playback format in an optimal manner under the assumed sound field model.

The framework is validated objectively in terms of reproduced binaural cue metrics for a correctly estimated as well as under- and over-estimated number of sources. Additionally, it is validated subjectively through a three-part binaural listening test using simulated Ambisonic receivers, SMAs, and head‑worn microphone arrays, with performance compared against binaural reference signals and a wide range of other state-of-the-art parametric rendering methods.

\section{Signal model}
\label{sec:signalModel}

\subsection{Preliminaries}

A direction-of-arrival (DOA) is denoted as $\u=[\cos\phi\cos\theta, \sin\phi\cos\theta, \sin\theta]^\rT$ for wave propagation at the origin incident from $(\phi,\theta)$ azimuth and elevation angles. Spherical integration over all directions is denoted as $ \sint{} = \int_{0}^{2\pi}\mathrm{d}\phi \int_0^{\pi} \sin\theta\mathrm{d}\theta$. Spherical harmonics (SHs) of order $n$ and degree $l$ are denoted as $Y_{nl}(\u)$ and are used throughout this work in their real orthonormalised form, compatible with the Ambisonics framework (ACN/N3D convention) \cite{driscoll1994computing}. Vectors of SHs up to a maximum order $N$ are denoted as $\y_N(\u) = [Y_{00}(\u), Y_{1(-1)}(\u), ...,Y_{nl}(\u),... Y_{NN}(\u)]^\rT$. Due to their orthonormality, the SHs satisfy
\begin{equation}
    \sint{\y_N^{}(\u)\y_N^\rT(\u)} = \hI_N, \label{eq:SHorthogonality}
\end{equation}
where $\hI_N$ is an $(N+1)^2\times(N+1)^2$ identity matrix.
The spherical harmonic transform (SHT) of a directional function $d(\u)$ up to order $N$ is given by
\begin{align}
    \hd_N^\rT = \mathcal{SHT}_N[d(\u)] = \sint{d(\u)\y_N^\rT(\u)},
\end{align}
returning SH coefficients $\hd_N$. For the rest of the manuscript, SHs and SH coefficients are indexed with $q=1,...,Q_N$, where $Q_N=(N+1)^2$, such that $Y_q\equiv Y_{nl}$ with $q=n^2+n+l+1$ and $\hd_N = [\hat{d}_1, ..., \hat{d}_{Q_N}]^\rT$. The directional function $d(\u)$ can then be recovered through the inverse SHT exactly, if it is spatially band limited to order $N$ or lower, with
\begin{align}
    d(\u) = \sum_{q=1}^{Q_N} \hat{d}_q Y_q(\u) = \hd_N^\rT \y_N^{}(\u).
    \label{eq:isht}
\end{align}

The SHT of a vector of $M$ directional functions $\d(\u) = [d_1(\u),...,d_M(\u)]^\rT$ is similarly defined as
\begin{align}
    \hD_N^{} = \mathcal{SHT}_N[\d(\u)] = \sint{\d(\u)\y_N^\rT(\u)},
\end{align}
where $\hD_N = [\hd_N^{(1)},...,\hd_N^{(M)}]^\rT$ is a $M\times Q_N$ matrix of SH coefficients stacked for all $d_m(\u)$ with $m=1,...,M$. The inverse SHT is then
\begin{align}
    \d(\u) = \hD_N \y_N(\u).
    \label{eq:misht}
\end{align}

A directional function $d(\u)$ rotated by a rotation matrix $\R$ results in the rotated $d^{(r)}(\u)$, such that $d^{(r)}(\u) = d(\R^{-1}\u)$. Equivalently, using the inverse SHT, this becomes
\begin{align}
    d^{(r)}(\u) &= d(\R^{-1}\u) = \hd_N^\rT\y_N(\R^{-1}\u) = \hd_N^\rT\hR_N^{-1}\y_N^{}(\u).
\end{align}
The $Q_N\times Q_N$ SHD rotation matrix $\hR_N$ can be computed from the standard Cartesian rotation matrix $\R$, e.g. as in \cite{ivanic1998rotation} and, since it is orthogonal, $\hR_N^{-1} = \hR_N^\rT$. Hence, the SH coefficients of the rotated function are
\begin{align} 
    \hd_N^{(r)} &= \hR_N^{} \hd_N^{}.
\end{align}
Similarly, for a vector of $M$ spherical functions $\d(\u)$ with SH coefficient matrix $\hD_N$ and a rotated version of those functions $\d^{(r)}(\u)$, its SH coefficient matrix is given as
\begin{align}
    \hD_N^{(r)} &= \hD_N^{} \hR_N^\rT . \label{eq:rotatedFuncs}
\end{align}

The integral of three SHs constitute a Gaunt coefficient
\begin{align}
    G_{q',q''}^q = \sint{Y_{q'}(\u)Y_{q''}(\u)Y_{q}(\u)}. 
\end{align}
Gaunt coefficients are useful for expressing the SH coefficients of a product of two spherical functions with respect to the SH coefficients of the two functions. They can be precomputed up to maximum orders $N'$, $N''$, $N$ of interest for both complex or real spherical SHs \cite{politis2024gaunt}. A more convenient form in this work is a matrix of Gaunt coefficients defined by the integral
\begin{align}
    \G_{N',N''}^q = \sint{\y_{N'}^{}(\u)\y_{N''}^\rT(\u)Y_{q}(\u)}. 
    \label{eq:Gauntmtx}
\end{align}

\subsection{Sound field model}

\begin{figure*}[!t]
\centerline{
\includegraphics[height=3.3cm]{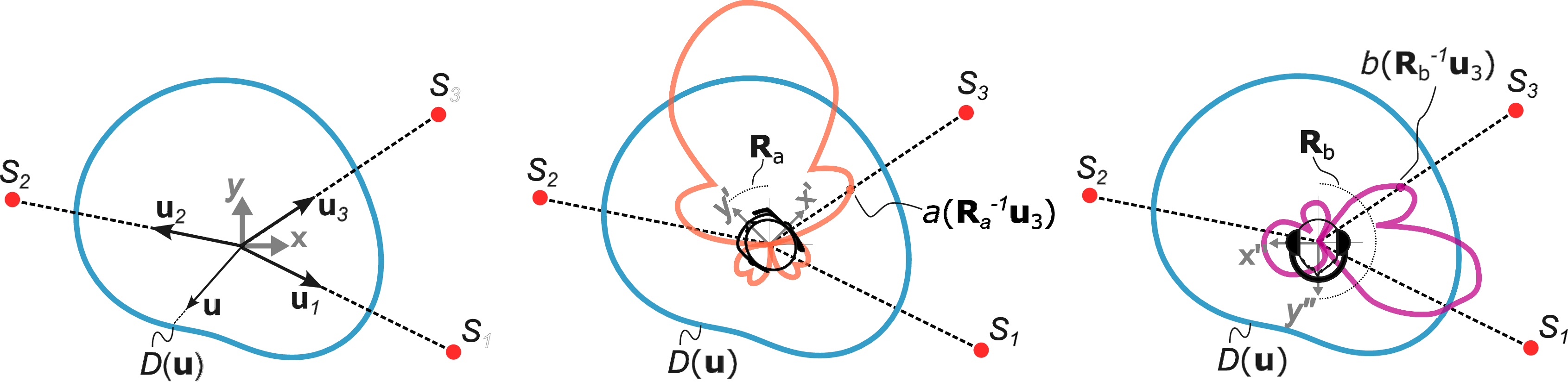}
}
\caption{The sound field model. The blue line indicates the ambience power distribution, while the orange and purple lines indicate the ATF of a single microphone in the capture array and the PTF of a single channel in the playback setup, respectively.}
\vspace{-0.3cm}
\label{fig:signal_model}
\end{figure*} 

First assume a sound scene comprising $k=1,...,K$ point-like sources in the far field, with DoAs $\u_k=[\cos\phi_k\cos\theta_k, \sin\phi_k\cos\theta_k, \sin\theta_k]^\rT$ at $(\phi_k,\theta_k)$ azimuth and elevation angles. Considering a time-frequency signal representation such as the short-time Fourier transform (STFT), with temporal and frequency indices $(t,f)$, the source waves carry signals $s_k(t,f)$ at the origin with power spectral densities $S_k(t,f) = \mathbb{E}[|s_k(t,f)|^2]$. 

In addition to the point sources, it is assumed that a general ambient sound field exists in the scene, which is directionally uncorrelated, carrying a signal $d(t,f, \u)$ from DoA $\u$ at the origin. Such an ambient sound field is characterised by the directional power distribution $D(t,f,\u)$ with the properties
\begin{align}
    \mathbb{E}[d(t,f, \u)d^*(t,f, \u')] &= \begin{cases}
  D(t,f,\u)  & \text{if  } \u = \u' \\
  0 & \text{if  } \u \neq \u',
\end{cases} \label{eq:dpow}\\
\mathbb{E}[d(t,f, \u)s_k^*(t,f)] &=
  0\quad\quad\quad \text{for  } k=1,...,K.
\end{align}

It is further assumed that the ambience sound field has a smooth power distribution $D$, which can be modelled by a finite number of SH coefficients up to order $N$ with $\hd_N^\rT(t,f) = \mathcal{SHT}_N[D(t,f,\u)]$.
The power distribution may then be recovered via the inverse SHT of Eq.~\ref{eq:isht}
\begin{equation}
    D(t,f,\u) = \sum_{q=1}^{Q_N} \hat{d}_q(t,f) Y_q(\u).
    \label{eq:shdpow}
\end{equation}

\subsection{Capture and reproduction signal model}

Consider an array around the origin capturing this sound scene, with $M$ recording channels. The array is defined generally, in the sense that it could comprise: spaced omnidirectional or directional sensors, microphones mounted on a spherical scatterer (i.e., SMAs \cite{rafaely2015fundamentals}), microphones mounted on an irregular scattering body such as a mobile phone \cite{delikaris2017parametric} or a head-mounted array \cite{donley2021easycom,fernandez2022enhancing}, a binaural dummy head or binaural recording of an individual using their own HRTFs, or an idealised Ambisonic microphone directly capturing SH coefficients of the sound field. 

In all cases, the directional array transfer functions (ATFs) are assumed to be known and given by $\a(f,\u) = [a_1(f,\u),...,a_M(f,\u)]^\rT$. 
Similarly a playback spatial audio format of $M'$ channels can be characterised by directional playback transfer functions (PTFs). The PTFs define how sound content from a certain direction is filtered and distributed to the playback channels. Typical cases of PTFs are, for example, ambisonic panning functions \cite{zotter2012all}, vector-base amplitude panning \cite{pulkki1997virtual} functions for loudspeaker rendering, or HRTFs for binaural playback on headphones. The PTFs are also assumed to be known and given by $\b(f,\u) = [b_1(f,\u),...,b_{M'}(f,\u)]^\rT$. 
 There is no fundamental distinction between ATFs and PTFs in this work; both are viewed as directional functions applied to the same sound field model that describes the spatial scene to be recorded or reproduced. 

Based on the above, the signals captured by the two systems, for the $K$ sources and ambience sound field, are given as
\begin{align}
    \x(t,f) &= \sum_k \a(f,\u_k)s_k(t,f)  + \sint{\a(f,\u) d(t,f,\u)},  \\
    \y(t,f) &= \sum_k \b(f,\u_k)s_k(t,f)  + \sint{\b(f,\u) d(t,f,\u)} .
\end{align}

The spatial properties of the sound field model are captured in the inter-channel signal statistics, expressed by their spatial covariance matrix (SCM). Assuming that the source and ambient signals are not correlated between them, these are
\begin{align}
    \X(t,f) = \mathbb{E}[\x(t,f)\x^\rH(t,f)] =  \X_s(t,f) + \X_d(t,f), \label{eq:recSCM} \\
    \Y(t,f) = \mathbb{E}[\y(t,f)\y^\rH(t,f)] =  \Y_s(t,f) + \Y_d(t,f), \label{eq:repSCM}
\end{align}
where $ \X_s$ and $\Y_s$ are the source-related SCMs, while $\X_d$ and $\Y_d$ denote the ambience-related SCMs. For the remaining presentation of the SCM model, only the recording case based on the ATFs $\a(f,\u)$ is formulated. The source related SCM is given as
\begin{align}
    \X_s(t,f) &= \sum_k S_k(t,f) \a(f,\u_k) \a^\rH(f,\u_k) \nonumber\\
    &= \sum_k S_k(t,f) \bA(f, \u_k),
    \label{eq:recsrcSCM}
\end{align}
where $\bA(f,\u) = \a(f,\u) \a^\rH(f,\u)$.

The SCM of the ambience signals can be expressed in terms of the ambience power distribution $D(t,f,\u)$
\begin{align}
    &\X_d(t,f) = \nonumber\\ & ~~~~ \Ex{\sint{\a(f,\u) d(t,f,\u)} \left(\sintt{\a(f,\u') d(t,f,\u')}\right)^\rH } \nonumber\\
    &=  \ssint{ \a(f,\u) \a^\rH(f,\u') \Ex{ d(t,f,\u)  d^*(t,f,\u')} }.
\end{align}
Using Eq.~\ref{eq:dpow}, the last relation reduces to
\begin{align}
    \X_d(t,f)
    &=  \sint{ \a(f,\u) \a^\rH(f,\u) D(t,f,\u) } \nonumber\\
    &= \sint{ \bA(f,\u) D(t,f,\u) }. \label{eq:recambSCM}
\end{align}
Note that modelling the SCM of the playback case $\Y(t,f)$ using PTFs $\b(f,\u)$ follows the same derivation, except substituting $\a$ and $\bA$ with $\b$ and $\bB$, respectively.

\subsection{Capture and playback under rotation}

Independent rotation of the ATF and the PTF are also of practical interest. Regarding ATFs, recording devices are often not aligned with the intended reference world coordinate frame or they can be dynamically rotating, as in the case of a head-mounted array \cite{stahl2024perceptual}. Regarding PTFs, rotation is necessary for head-tracked binaural playback in devices such as headphones or head-mounted displays \cite{mccormack2023six}. Considering a potential rotation $\R_a$ for the recording setup and a potential rotation $\R_b$ for the playback setup, the signals captured by the two systems for the $K$ sources and ambience sound field are
\begin{align}
    \x(t,f) =& \sum_k \a(f,\R_a^{-1}\u_k)s_k(t,f) \nonumber + \\& \sint{\a(f,\R_a^{-1}\u) d(t,f,\u)}, \\  
    \y(t,f) =& \sum_k \b(f,\R_b^{-1}\u_k)s_k(t,f) \nonumber + \\& \sint{\b(f,\R_b^{-1}\u) d(t,f,\u)}.
\end{align}
Following Eq.~\ref{eq:recsrcSCM} \& \ref{eq:recambSCM}, the capture SCMs are therefore
\begin{align}
    \X_s(t,f)
    &= \sum_k S_k(t,f) \bA(f, \R_a^{-1}\u_k),\\
 \X_d(t,f)
    &=  \sint{ \bA(f,\R_a^{-1} \u) D(t,f,\u) }. \label{eq:recrotSCM}
\end{align}
The respective playback SCMs $\Y_s, \Y_d$ are defined similarly.



\section{Spatial rendering approaches}
\label{sec:renderingMethods}

The objective of a spatial rendering method is to transform the $M$-channel signals $\x$ captured by a recording system, into signals $\y$ that approximate those the PTFs of an $M'$-channel reproduction system would have recorded had they been used to capture the same acoustic scene.
A distinction can also be made between signal-independent non-parametric solutions and signal-dependent parametric alternatives. Non-parametric approaches depend only on the recording and playback setup properties as characterised by the ATFs and PTFs
\begin{equation}
    \y(t,f) = \T(f)\x(t,f),
\end{equation}
where $\mathbf{T}$ is a $M' \times M$ matrix of real or complex gains. Typically, such matrices are optimised jointly for all possible directions on the sphere and are often time-invariant. For cases where the capture resolution is low, the error for any given direction can be high; resulting in ambiguous localisation cues and/or unnatural sounding reproduction of diffuse ambience \cite{bertet2013investigation, avni2013spatial}.

Parametric signal‑dependent approaches differ, in that they rely not only on the input signals but also on estimated sound field model parameters $\tilde{\p}(t,f)$. These parameters are extracted during analysis and then used during synthesis to produce time‑varying mixing matrices
\begin{equation}
    \y(t,f) = \M(t,f|\tilde{\p}(t,f))\x(t,f), \label{eq:sigDepModel}
\end{equation}
which can be optimised with respect to perceptual attributes; allowing the reproduced scene to be perceptually closer to the target even when the minimisation objective (e.g., matching SCMs, spatial statistics, interchannel level differences, or correlations) is not perfectly satisfied. Typically, such methods optimise the linear mapping for a small set of dominant directions, guided by time-frequency‑dependent source DOA estimates.

In the remainder of Sec.~\ref{sec:renderingMethods}, some popular non-parametric and parametric methods, which serve as comparison methods against the proposed framework, are presented in brief.

\subsection{{Least squares solution (LS)}}
A straightforward way to realize the spatial rendering objective is to seek a signal-independent $M'\times M$ transformation matrix $\T$ that would transform the ATFs of the recording setup $\a(\u)$ to the PTFs of the playback setup, $\b(\u)$, in a least squares sense over all directions
\begin{align}
    \argmin_\T \sint{ ||\T(f)\a(f,\u) - \b(f,\u) ||^2 }.\label{eq:leastsquares}
\end{align}
The same objective can be expressed more conveniently in terms of a spherical harmonic representation of the ATFs and PTFs (SH-ATFs and SH-PTFs). It is first assumed that the capture ATFs are band-limited to a suitable order $L$, with $\hA_L(f) = \mathcal{SHT}_L[\a(f,\u)]$ and, similarly, the PTFs up to a suitable order $L'$, with $\hB_{L'}(f) = \mathcal{SHT}_{L'}[\b(f,\u)]$. Both can then be expressed through the inverse SHT of Eq.~\ref{eq:isht} as
\begin{align}
    \a(f,\u) &= \hA_L(f)\y_L^{}(\u), \label{eq:shatf} \\
    \b(f,\u) &= \hB_{L'}^{}(f)\y_{L'}^{}(\u). \label{eq:shptf}
\end{align}
Using the above relations, the minimisation objective of Eq.~\ref{eq:leastsquares} becomes
\begin{align}
\label{eq:shleastsquares}
\argmin_\T \sint{ ||\T(f)\hA_{L}^{}(f)\y_{L}^{}(\u)
    - \hB_{L'}^{}(f)\y_{L'}^{}(\u) ||^2 }.
\end{align}
In cases where the capture setup is spatially more complex than the reproduction setup, with $L>L'$, Eq.~\ref{eq:shleastsquares} can be rewritten as
\begin{align}
\argmin_\T \sint{ ||\T(f)\hA_{L}^{}(f)\y_{L}^{}(\u)
    - \hB_{L'}^{}(f)\hI_{L',L}^{}\y_{L}^{}(\u) ||^2 },
\end{align}
where $\hI_{L',L} = [\hI_{L'}^{},\quad \mathbf{0}_{Q_{L'}\times (Q_L-Q_{L'})}]$ is a zero padded identity matrix extracting the $L'$th lower order components from an $L$th order vector of SH values. 

Similarly, in the case where the reproduction setup is spatially more complex than the capture setup, with $L< L'$, Eq.~\ref{eq:shleastsquares} can be rewritten as
\begin{align}
\argmin_\T \sint{ ||\T(f)\hA_{L}^{}(f)\hI_{L,L'}^{}\y_{L'}^{}(\u)
    - \hB_{L'}^{}(f)\y_{L'}(\u) ||^2 }.
\end{align}

Using the matrix trace properties $\x^\rH\x=\Tr{\x\x^\rH}$ and $\Tr{\X^\rH\X} = ||\X||_\rF^2$ and the SH property of Eq.~\ref{eq:SHorthogonality} the objective can be brought into the standard least squares form 
\begin{align}
\label{eq:lsobjective}
\argmin_\T ||\T(f)\hA_{L}^{}(f) -
     \hB_{L'}^{}(f)\hI_{L',L}^{}||^2_\rF,\quad \mathrm{for}\, L>L' & \nonumber\\
\argmin_\T ||\T(f)\hA_{L}^{}(f)\hI_{L,L'}^{} -
     \hB_{L'}^{}(f)||^2_\rF,\quad \mathrm{for}\, L<L' & .  
\end{align}

Regularisation is often added to the solution to control excessive amplification in the inversion, arising from limitations of the capture setup matching the playback setup.
By adding a regularisation term $\beta^2||\T(f)||^2_\rF$ to the minimisation objective, the regularised least squares solution of Eq.~\ref{eq:lsobjective} finally becomes
\begin{align}
   & \T(f) = \nonumber\\
   & \begin{cases}
    \hB_{L}^{}(f) \hA_{L}^\rH(f)\left(\hA_{L}^{}(f) \hA_{L}^\rH(f) + \beta^2\I_M  \right)^{-1} & \text{if  } L\leq L' \\
    \hB_{L'}^{}(f) \hA_{L'}^\rH(f)\left(\hA_{L}^{}(f) \hA_{L}^\rH(f) + \beta^2\I_M  \right)^{-1}  & \text{if  } L > L'.
\end{cases}\label{eq:shleastsquares_reg}
\end{align}


\subsection{{Magnitude least squares solution (MagLS)}}

In the case of headphone reproduction (i.e., binaural rendering), where the PTFs correspond to HRTFs, the least‑squares solution of Eq.~\ref{eq:shleastsquares_reg} is unable to approximate the spatially complex mid‑ and high‑frequency structure of the HRTFs when only small microphone arrays (or low‑order Ambisonic signals) are available \cite{schorkhuber2018binaural}. Based on the observation that at mid-high frequencies inter-aural level differences (ILDs) dominate auditory localisation, it may be preferable to shift the effort of approximating the complex HRTFs, to instead approximating only their magnitudes. This often leads to significantly improved binaural rendering over an extended frequency range, compared to the baseline LS solution. It is formulated as
\begin{align}
    \argmin_\T \lambda(f)\sint{ ||\T(f)\a(f,\u) - \b(f,\u) ||^2 }\, + \nonumber\\
    (1-\lambda(f))\sint{ ||\, |\T(f)\a(f,\u)| - |\b(f,\u)|\, ||^2 },
\end{align}
where 
\begin{align}
    \lambda(f) &= \begin{cases}
  1  & \text{if  } f < f_\rT \\
  0 & \text{if  } f \geq f_\rT.
\end{cases}
\end{align}

The solution applies standard LS of Eq.~\ref{eq:lsobjective} in the range up to the transition frequency index $f_\rT$ where the HRTFs are directionally smooth enough to be approximated accurately by the capture ATFs, including the important interaural phase- and time-difference (IPD/ITD) reconstruction for low-frequency localisation. Above $f_\rT$, the solution shifts to magnitude LS (MagLS), ignoring reconstruction of IPDs/ITDs, which become perceptually less relevant. The MagLS solution was formulated originally for ambisonic signals, i.e. $\a(f,\u)=\y_N(\u)$, in \cite{schorkhuber2018binaural}, but was later applied to head-mounted and other ATFs \cite{deppisch2021end,stahl2024perceptual}. For details regarding a practical implementation of the solution, the reader is referred to \cite{zotter2019ambisonics}.

\subsection{{Linearly \& quadratically constrained solution (LQCLS)}}
In \cite{schorkhuber2019linearly}, signal-dependent constraints were added to the otherwise signal-independent LS solution of Eq.~\ref{eq:leastsquares}, under the assumptions of a single source model ($K=1$) and an isotropic diffuse ambience component ($d_{q>1}=0$).
 The formulation simplifies to a linear distortionless constraint on the source DOA and a quadratic signal-independent constraint matching the diffuse field coherence of the ATFs to that of the PTFs
 \begin{align}
 \label{eq:lqclsobjective}
    & \argmin_\T \sint{ ||\T(f)\a(f,\u) - \b(f,\u) ||^2 },\quad \mathrm{s.t.} \nonumber\\
    & \T(f)\a_s(t,f) = \b_s(t,f), \nonumber\\
    & \T(f)\hA_L^{}(f)\hA_L^\rH(f)\T^\rH(f) = \hB_{L'}^{}(f)\hB_{L'}^\rH(f).
\end{align}
 where $\a_s, \b_s$ are the ATFs/PTFs for the analyzed source DOA.
 The solution can be elegantly parameterised w.r.t. only the source DOA.
 Note that whether the constraints can be met, in order for the solution of Eq.~\ref{eq:lqclsobjective} to be tractable, depends on properties of the PTFs; therefore, the authors in \cite{giller2019super} proposed to pre-process the target HRTFs, in order to ensure a stable solution for all DOAs. 

\subsection{{Directional Audio Coding (DirAC)}}

DirAC was one of the first signal-dependent parametric rendering methods. While originally designed for content captured in the first-order Ambisonics format \cite{pulkki2017first}, it was later generalised to higher-order Ambisonics input \cite{politis2015sector}. Its model has been additionally adapted to an additional number of specialised capture formats, such as spaced microphones for music recording \cite{politis2015parametric}, linear arrays \cite{thiergart2011parametric}, and head-worn arrays \cite{fernandez2022enhancing}. The original formulation, however, relied on specific properties of FOA signals, which are briefly described here. DirAC estimates three sound field parameters per time-frequency tile or subband window: the diffuseness $\psi$, the energy density $E$, and the DoA $\u$ obtained through acoustic intensity vector analysis. It then (i) applies a signal-independent rendering for the playback setup, using e.g. an ambisonic decoding matrix $\T$, (ii) applies a time-frequency soft mask based on diffuseness to realise a primary-ambience separation of the rendered signals, (iii) sharpens the primary component based on PTFs and the analysed DoA, and (iv) forces diffuse properties onto the ambience component via decorrelation operations. For example, for loudspeaker rendering to $M'$ loudspeakers with frequency-independent PTFs $\b(\u)$ given by VBAP gains, together with a frequency-independent ambisonic decoding matrix $\T$ (e.g., \cite{zotter2012all}), this becomes
\begin{align}
\label{eq:diracmixing}
    \z_s(t,f) &= \sqrt{1-\psi(t,f)} \T\x(t,f), \nonumber\\
    \z_a(t,f) &= \sqrt{\psi(t,f)} \T\x(t,f), \nonumber\\
    \y(t,f) &= \b_s(t,f) \circ \z_s(t,f) + \mathcal{D}[\z_a(t,f)],
\end{align}
where $\circ$ denotes element-wise multiplication between two vectors, and $\mathcal{D}[\cdot]$ denotes the application of independent decorrelators to each signal in the input signal vector. An example of suitable multi-channel signal decorrelation includes: the combination of fixed sub-band-dependent delays (with longer delayes at lower frequencies) and cascaded lattice all-pass filters (with higher orders at lower frequencies), followed by envelope shaping; such as those used within certain multi-channel audio codecs \cite{herre2008mpeg}. 

Recent implementations of the method, including higher-order DirAC \cite{politis2015sector}, avoid directly defining the rendering matrices as in Eq.~\ref{eq:diracmixing}. Instead they utilise an optimal mixing framework, which is also leveraged by the proposed method and is detailed in Sec.~\ref{sec:optimalmixing}. To that purpose, a modelled covariance matrix of the output channels $\Y= \Y_s+\Y_d$ must be specified. Optimal-mixing DirAC does this based on the analysed parameters as
\begin{align}
    \Y_s(t,f) &= (1-\psi(t,f))E(t,f) \b_s^{}(t,f)\b_s^\rH(t,f), \nonumber\\
    \Y_d(t,f) &= \psi(t,f)E(t,f)\boldsymbol{\Gamma}(f),
\end{align}
where $\boldsymbol{\Gamma}$ is a suitable diffuse field coherence matrix for the playback setup \cite{politis2016diffuse}. This variant of DirAC is the one implemented as one of the comparison methods in this study.

\subsection{{DOA-dependent spatial filtering solution (COMPASS)}}
The COMPASS method, as described in \cite{politis2018compass}, uses only the analysed source DOAs to design a $K\times M$ matrix $\W_s$ of spatial filters with a distortionless constraint towards each DOA, while placing nulls to the other DOAs if possible. These source signals are then spatialised over the playback format, based also on these analysed DOAs. An ambience residual is also computed at the recording format through an $M\times M$ blocking matrix $\W_a$. The resulting ambience signals are further spatially decoded to a small set of $V$ surrounding virtual sources through another $V\times M$ beamforming matrix $\W_\mathrm{vs}$ with energy preserving properties. The directional incoherence assumption of the ambience sound field is enforced, by subjecting these virtual source signals to decorrelation operations. Finally, the decorrelated signals are spatialised over the target playback format in their respective directions. The process can be summarised as
\begin{align}
    \y(t,f) &= \M_s(t,f)\x(t,f) + \B_\mathrm{vs}(f)\mathcal{D}[\M_a(t,f)\x(t,f)], \nonumber\\
    &\mathrm{with}\nonumber\\
    \M_s(t,f) &= \B_s(t,f)\W_s(t,f) \nonumber\\
    \M_a(t,f) &= \W_\mathrm{vs}(f)\left(\I_M-\A_s(t,f)\W_s(t,f)\right),
\end{align}
 where $\A_s(f)= [\a(f,\u_1),.,,,\a(f,\u_K)]$ is the matrix of ATFs for the source DOAs, $\B_s(f)$ is defined accordingly for the PTFs, while $\B_\mathrm{vs}(f)$ is the matrix of PTFs for the directions of the virtual sources. Rather than decorrelating ambient signals through virtual sources, a more efficient and robust variation that minimises decorrelator usage is described in \cite{mccormack2022estimating}. This variant was also included as one of the comparison methods.




\begin{figure*}[!t]
\centerline{
\includegraphics[width=16.2cm,trim={0.9cm 19.8cm 1.4cm 0.79cm}, clip]{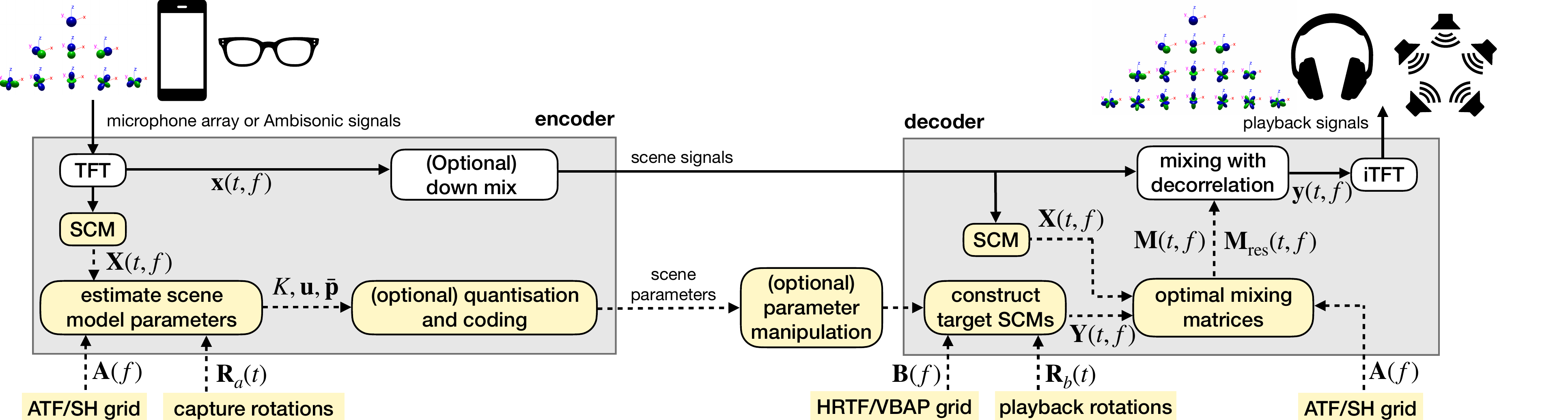}
}
\caption{Block diagram of the proposed framework.}
\vspace{-0.3cm}
\label{fig:block_diagram}
\end{figure*} 

\section{Proposed rendering framework ({\normalfont n}COMPASS)}
\label{sec:proposed}

The proposed nCOMPASS framework performs spatial rendering by combining a detailed sound‑field analysis with an optimal mixing formulation that adapts to both the capture and playback formats. The framework is format‑agnostic. It accepts either raw microphone‑array signals or Ambisonic signals as input, handles rotations at both capture and reproduction, and supports rendering to arbitrary playback formats, including loudspeaker‑based reproduction, binaural output, or re‑encoding to Ambisonics.

As illustrated in Fig.~\ref{fig:block_diagram}, the method consists of two main stages. In the \emph{decoding} stage, the input signals are rendered using adaptive time-frequency mixing matrices, as described in Eq.~\ref{eq:sigDepModel}. Computing these mixing matrices requires knowledge of the output signal SCMs $\Y(t,f)$, which are not directly available. Instead, they are constructed based on the assumed sound‑field model of Sec.~\ref{sec:signalModel}, comprising a variable number of point/plane‑wave sources, together with a directional ambience component. The \emph{encoding} stage estimates the sound‑field parameters required to construct these SCMs. These parameters, transmitted to the decoder alongside the input signals (optionally after compression or quantisation), are the time-frequency‑dependent number of sources $K(t,f)$, their DOAs $\u_k(t,f)$, their powers $S_k(t,f)$, and the $Q_N(t,f)$ SH coefficients $\hat{d}_q(t,f)$ describing the ambience power distribution. A detailed description of both stages follows.


\subsection{nCOMPASS decoding \& optimal mixing}
\label{sec:optimalmixing}

nCOMPASS employs the generic and flexible rendering solution proposed by Vilkamo et al.~\cite{vilkamo2013optimized}, which has also been adopted by the present authors for higher‑order extensions of DirAC~\cite{politis2015sector,politis2017enhancement} and for optimised ambience rendering in COMPASS~\cite{mccormack2022estimating}. The mixing solution $\M$ is fully signal‑dependent. It enforces a quadratic constraint that matches the full target playback SCM, while simultaneously minimising, in a least‑squares sense, a signal‑fidelity error term derived from a predefined signal‑independent mixing matrix $\T$, obtained for example through the least‑squares solution of Eq.~\ref{eq:shleastsquares_reg}. The constraint ensures accurate spatial rendering, whereas the error‑minimisation term promotes high signal fidelity and suppresses time-frequency processing artefacts. It can be formulated as
\begin{align}
    & \argmin_\M \Ex{||\M(t,f)\x(t,f) - \tT(f)\x(t,f) ||^2 },\quad \mathrm{s.t.} \nonumber\\
    & \M(t,f)\X(t,f)\M^\rH(t,f) = \Y(t,f). \label{eq:optimalMixing}
\end{align}
The matrix $\tT$ equals the provided prototype decoding matrix $\T$ after normalisation to produce signals that match the target channel energies in $\Y$; i.e. $\tT = \mathbf{\Xi}\T$ where $\mathbf{\Xi}$ is a diagonal matrix with entries $[\mathbf{\Xi}]_{ii} = \sqrt{[\Y]_{ii}/[\T\X\T^\rH]_{ii}}$.

Matrices that satisfy the quadratic constraint of Eq.~\ref{eq:optimalMixing} can be constructed through the eigendecomposition of the SCMs involved, i.e. $\X = \V_\x^{}\E_\x^{}\V_\x^\rH = \K_\x^{}\K_\x^\rH$ with $\K_\x^{}=\V_\x^{}\E_\x^{1/2}$. Similarly, $\Y = \K_\y^{}\K_\y^\rH$. Assuming for simplicity that the number of recording and playback channels is the same $M=M'$, the solution is parameterised as
\begin{equation}  
\M(t,f) = \K_\y^{}(t,f)\U(t,f)\K_\x^{-1}(t,f),\label{eq:OMsolutionM}
\end{equation}
for an arbitrary unitary matrix $\U$. Inserting this parameterised solution into Eq.~\ref{eq:optimalMixing} and using the properties $\z = \K_\x^{-1}\x$ and $\Ex{\z\z^\rH} = \I_M$, an equivalent minimisation problem with a unitary constraint can be formulated as
 \begin{align}
    & \argmin_\U ||\K_\y(t,f)\U(t,f) - \tT(f)\K_\x(t,f)||_\rF^2,\quad \mathrm{s.t.} \nonumber\\
    & \U^\rH(t,f)\U(t,f) = \I_M.
\end{align}
This is a unitary Procrustes problem with the well known solution
\begin{equation}
	\U(t,f) = \V(t,f)\Z^\rH(t,f),
\end{equation}
where $\V$ and $\Z$ are the right and left singular vectors obtained from the singular value decomposition (SVD) $\Z\boldsymbol{\Sigma}\V^\rH = \K_\x^\rH \tT^\rH \K_\y$. In cases where the number of recording and playback channels differ, $M\neq M'$, the solution is modified to
\begin{equation}
	\U(t,f) = \V(t,f)\tI_{M', M} \Z^\rH,\label{eq:OMsolutionU}
\end{equation}
where $\tI$ is a zero-padded $M'\times M$ identity matrix matching the number of output channels to the number of input channels, in cases where the optimal mixing is on upmixing $M'>M$ or downmixing $M'<M$ duties. Using the solution of Eq.~\ref{eq:OMsolutionU} in Eq.~\ref{eq:OMsolutionM} provides the optimal mixing matrix for the nCOMPASS rendering objective.

As \cite{vilkamo2013optimized} notes, a robust solution that meets the target SCM purely by mixing is not always possible, expressed by an unstable $\K_\x^{-1}$ in Eq.~\ref{eq:OMsolutionM} requiring regularisation. However, the target SCM can always be met if decorrelated signal energy is injected into the original mixing solution, similar to how the other mentioned rendering solutions (and upmixing in general) utilise decorrelation to restore diffuse ambience over the target playback system. This can be achieved in an optimal manner, which minimises the amount of decorrelation required, as
\begin{equation}
        \y(t,f) = \M(t,f)\x(t,f) + \M_\mathrm{res}(t,f)\mathcal{D}[\T(f)\x(t,f)],
\end{equation}
where $\M$ is the primary mixing solution, while $\M_\mathrm{res}$ is the residual $M'\times M'$ mixing solution that mixes in decorrelated versions of the prototype signals $\T\x$. The residual mixing matrix can be derived in a similar manner to the primary mixing matrix, but with the residual SCM constraint $\M_\mathrm{res}^{}\Y_\mathrm{p}\M_\mathrm{res}^\rH = \Y - \M\X\M^\rH$. Note that $\Y_\mathrm{p}=\mathrm{diag}\left[ \T\X\T^\rH \right]$ is the diagonal SCM of the decorrelated prototype signals to be mixed, serving as input to the residual mixing stage. For more details on the derivation of the solution, the reader is referred to \cite{vilkamo2013optimized}. 

\subsection{nCOMPASS encoding \& estimation of parameters}
\label{sec:estModelParams}

In order to construct the target SCM required for the rendering solution, the parameters of the sound field model in Sec.~\ref{sec:signalModel} must first be estimated. The narrowband detection of the number of active sources $K$ and the estimation of their DOAs $\u_k$ is outside the scope of this work; the proposed framework instead simply adopts the same subspace-based estimators used in COMPASS \cite{politis2018compass}. The focus of this section is therefore on estimating the source powers $S_k$ and the ambience power coefficients $\hd_N$, while also accounting for potential rotations $\R_a$ and $\R_b$ of the capture and reproduction setups.

Ignoring rotations initially, and using Eqs.~\ref{eq:recsrcSCM} and \ref{eq:recambSCM}, the capture SCM can be written as
\begin{equation}
    \X(t,f) = \sum_k S_k(t,f) \bA(f, \u_k) + \sint{ \bA(f,\u) D(t,f,\u) }.
\end{equation}
To express the ambience term in a more tractable form, the SH expansions of the ATFs (Eq.~\ref{eq:shatf}) and of the ambience power distribution (Eq.~\ref{eq:shdpow}) are substituted. This yields
\begin{align}
   \X_d(t,f) =
      \sint{ \hA_L(f)\y_L(\u) \y_L^\rT(\u) \hA_L^\rH(f) D(t,f,\u) } \nonumber\\
    =  \sint{ \hA_L(f)\bY_L(\u) \hA_L^\rH(f)
        \left(\sum_{q=1}^{Q_N} \hat{d}_q(t,f)Y_q(\u)\right)} \nonumber\\
    =  \sum_{q=1}^{Q_N} \hat{d}_q(t,f)
        \left(\hA_L(f) \left(\sint{ \bY_L(\u) Y_q(\u)}\right) \hA_L^\rH(f)\right),
\end{align}
where $\bY_L(\u)=\y_L(\u)\y_L^\rT(\u)$. The integral term depends only on the SH order $L$ and forms a matrix of Gaunt coefficients (Eq.~\ref{eq:Gauntmtx})
\begin{align}
    \G_L^q = \sint{ \bY_L(\u)Y_q(\u)}.
\end{align}
Thus, the ambience SCM becomes
\begin{align}
    \X_d(t,f)
    &= \sum_{q=1}^{Q_N} \hat{d}_q(t,f)
       \left(\hA_L(f)\G_L^q\hA_L^\rH(f)\right) \nonumber\\
    &= \sum_{q=1}^{Q_N} \hat{d}_q(t,f)\H_L^q(f),
    \label{eq:SHrecambSCM}
\end{align}
where the matrices $\H_L^q(f)=\hA_L(f)\G_L^q\hA_L^\rH(f)$ are scene-independent and can be precomputed.




When the capture array undergoes a rotation $\R_a$, the ambience SCM can be written as
\begin{align}
    \X_d(t,f) &=  \sum_{q=1}^{Q_N} \hat{d}_q(t,f) \H_L^q(f,\R_a),\;\mathrm{with} \\
    \H_L^q(f,\R_a) &= \hA_L^{}(f) \hR_L^\rT \G_L^q \hR_L^{} \hA_L^\rH(f).
\end{align}
However, updating $\H_L^q(f,\R_a)$ continuously for a rotating capture setup, or equivalently playback setup (e.g., as in the case of head-tracked headphones), would be computationally expensive. Therefore, rather than rotating the ATFs, the same result can be achieved if the ambience power distribution is rotated in the inverse manner
\begin{align}
    \X_d(t,f)
    =  \sint{ \bA(f, \u) D(t,f,\R_a \u) }. \label{eq:invRotAmbSCM}
\end{align}
Following previous derivations, Eq.~\ref{eq:invRotAmbSCM} becomes
\begin{align}
    \X_d(t,f) =  \sum_{q=1}^{Q_N} \hat{d}_q^{(\rr)}(t,f) \H^q_L(f),
\end{align}
where $\hat{d}_q^{(\rr)}$ is the $q$th rotated coefficient of $\hd_N^{(\mathrm{r})} = \hR_N^\rT \hd_N$.

The complete SCMs models for the capture side and playback side under rotation are therefore
\begin{align}
    \X(t,f) &= \sum_{k=1}^K S_k(t,f) \bA(f, \R_a^{-1}\u_k) +  \sum_{q=1}^{Q_N} \hat{d}_q^{(\mathrm{r})}(t,f) \H_L^q(f), \\
     \Y(t,f) &= \sum_{k=1}^K S_k(t,f) \bB(f, \R_b^{-1}\u_k) +  \sum_{q=1}^{Q_N} \hat{d}_q^{(\mathrm{r})}(t,f) \F_{L'}^q(f), 
\end{align}
where $\bB(f,\u)=\b(f,\u)\b^\rH(f,\u)$ and
$\F_{L'}^q(f)=\hB_{L'}(f)\G_{L'}^q\hB_{L'}^\rH(f)$ are defined analogously to $\bA$ and $\H_L^q$.

A linear estimator for the model parameters can now be constructed \cite{friedlander1995direction}. The source SCM can be expressed in a vectorised form $\bx_s = \mathrm{vec}[\X_s]$ as
\begin{align}
    \bx_s(t,f) &= \sum_{k=1}^K S_k(t,f) \ba(f, \R_a^{-1}\u_k) \nonumber\\
    &= \C^{(\rr)}(f) \p_s(t,f),
\end{align}
with 
\begin{align}
    \ba(f, \u) &= \mathrm{vec}[\bA(f, \u)] \label{eq:vecsrcATF},\\
    \C^{(\rr)}(f) &= [\ba(f, \R_a^{-1}\u_1),..., \ba(f, \R_a^{-1}\u_K)] \label{eq:vecsrcMTX},\\
    \p_s(t,f) &= [S_1(t,f),...,S_K(t,f)]. 
\end{align}
Similarly, vectorising the ambience SCM $\bx_a = \mathrm{vec}[\X_a]$ yields
\begin{align}
    \bx_a(t,f) &=  \sum_{q=1}^{Q_N} \hat{d}_q^{(\rr)}(t,f) \bh^q_L(f) = \L(f) \hd_N^{(\rr)}(t,f) \nonumber\\
    &= \L(f) \hR_N^\rT \hd_N^{}(t,f) = \L^{(\rr)}(f) \hd_N(t,f),
\end{align}
where 
\begin{align}
    \bh^q_L(f) &= \mathrm{vec}[\H^q_L(f)], \label{eq:vecambATF}\\
    \L(f) &= [\bh^1_L(f),..., \bh_L^{Q_N}(f)], \\
    \L^{(\rr)}(f) &= \L(f) \hR^\rT_N. \label{eq:vecambMTX}
\end{align}

Combining the above, the vectorised array SCM $\bx = \mathrm{vec}[\X]$ can be expressed as 
\begin{align}
    \bx(t,f) = \E(f) \p(t,f),
\end{align}
where $\E = [\C^{(\rr)}, \L^{(\rr)}]$ and $\p = [S_1, ..., S_K, \hat{d}_1, ..., \hat{d}_{Q_N}]^\rT$ is the vector of $K+(N+1)^2$ source and ambience power parameters.

Given $K$, the DOAs $\u_k$, and known ambience power distribution order, the power parameters are estimated via
\begin{align}
    \tilde{\p}(t,f) = \E^+(f)  \bx(t,f),
\end{align}
where $()^+$ denotes the Moore-Penrose pseudoinverse. Note that due to the Hermitian property of the matrices involved, the estimates are guaranteed to be real valued.

The vectorised playback SCM is then obtained from
\begin{align}
    \by(t,f) = \D(f) \p(t,f),
\end{align}
where $\D=[\N^{(\mathrm{r})},\O^{(\mathrm{r})}]$ is constructed analogously to $\E$, using $\bB$ and $\F_{L'}^q$ in place of $\bA$ and $\H_L^q$. The matrix $\Y$ is finally recovered as $\Y=\mathrm{vec}^{-1}[\by]$, where $\mathrm{vec}^{-1}[\cdot]$ reconstructs the original matrix from its vectorised version.

The number of scene parameters $K$ and $Q_N$ that can be estimated depends primarily on the number of microphones/capture channels $M$. Due to the Hermitian structure of the matrices involved, the maximum number of real parameters that can be estimated is $K+Q_N = M^2$. In the absence of any detected source components ($K=0$), the ambience power distribution can be modelled up to a maximum SH order of $N = M - 1$. In typical narrowband detection and DOA estimation methods employed in parametric spatial audio processing, the number of reliably localised sources is strictly less than the number of microphones ($K<M$). For example, if a maximum of $K = \lfloor M/2 \rfloor$ sources are detected and localised, then the maximum ambience SH order that can be estimated is $N=M-2$. Hence, for a small number of assumed or detected source components, the ambience power distribution becomes increasingly spatially complex with the number of microphones. This allows the ambience model to compensate for a potential underestimation of primary source components and to account for the presence of weaker or spread directional components in a spatially complex scene.

\section{Binaural Metrics Analysis}
\label{sec:objective}

\begin{table*}[t!]
\centering
\caption{Timbral colouration (dB) RMSE values.}
\begin{tabular}{ccc| *{3}{ccc}}
\toprule
\multicolumn{3}{c|}{\textbf{Input sound scene}} 
    & \multicolumn{3}{c}{\textbf{$K_\mathrm{est}=0$}} 
    & \multicolumn{3}{c}{\textbf{$K_\mathrm{est}=1$}} 
    & \multicolumn{3}{c}{\textbf{$K_\mathrm{est}=2$}} \\
\cmidrule(lr){1-3}
\cmidrule(lr){4-6}
\cmidrule(lr){7-9}
\cmidrule(lr){10-12}
{\scriptsize \textbf{$K_\mathrm{gt}$}} & {\scriptsize \textbf{$N_\mathrm{d}$}} & {\scriptsize {SAR} (dB)}
    & {\scriptsize \textit{MagLS}} & {\scriptsize \textit{COMPASS}} & {\scriptsize \textit{nCOMPASS}}
    & {\scriptsize \textit{MagLS}} & {\scriptsize \textit{COMPASS}} & {\scriptsize \textit{nCOMPASS}}
    & {\scriptsize \textit{MagLS}} & {\scriptsize \textit{COMPASS}} & {\scriptsize \textit{nCOMPASS}} \\
\midrule
0 & 0 & $-\infty$
    & \grey \textbf{0.0302} & \grey 0.3033 & \grey 0.0340
    & \textbf{0.0302} & 0.4507 & 0.0349
    & \textbf{0.0302} & 1.1163 & 0.0476 \\
0 & 1 & $-\infty$
    & \grey 0.0457 & \grey 0.2634 & \grey \textbf{0.0325}
    & 0.0457 & 0.4484 & \textbf{0.0327}
    & \textbf{0.0457} & 1.2875 & 0.0603 \\
1 & N/A & $\infty$
    & \textbf{2.4577} & 2.8063 & 2.7948
    & \grey 2.4577 & \grey \textbf{0.0000} & \grey \textbf{0.0000}
    & 2.4577 & \textbf{0.0000} & \textbf{0.0000} \\
1 & 0 & 0
    & \textbf{0.7937} & 0.8783 & 0.8635
    & \grey 0.7937 & \grey 0.2297 & \grey \textbf{0.0311}
    & 0.7937 & 0.8837 & \textbf{0.0322} \\
1 & 1 & 0
    & \textbf{0.7640} & 0.8608 & 0.8470
    & \grey 0.7640 & \grey 0.2335 & \grey \textbf{0.0286}
    & 0.7640 & 0.9362 & \textbf{0.0324} \\
2 & N/A & $\infty$
    & \textbf{1.4164} & 1.5488 & 1.5044
    & 1.4164 & 1.1208 & \textbf{1.1810}
    & \grey 1.4164 & \grey \textbf{0.0178} & \grey 0.0285 \\
2 & 0 & 0
    & \textbf{0.5789} & 0.6829 & 0.6271
    & 0.5789 & 0.5234 & \textbf{0.4483}
    & \grey 0.5789 & \grey 0.8319 & \grey \textbf{0.0310} \\
2 & 1 & 0
    & \textbf{0.5497} & 0.6659 & 0.6040
    & 0.5497 & 0.5527 & \textbf{0.4344}
    & \grey 0.5497 & \grey 0.7854 & \grey \textbf{0.0329} \\
\bottomrule
\end{tabular}
\label{table:timbre_rmse} 
\end{table*}

\begin{table*}[t!]
\centering
\caption{ILD (dB) RMSE values.}
\begin{tabular}{ccc| *{3}{ccc}}
\toprule
\multicolumn{3}{c|}{\textbf{Input sound scene}} 
    & \multicolumn{3}{c}{\textbf{$K_\mathrm{est}=0$}} 
    & \multicolumn{3}{c}{\textbf{$K_\mathrm{est}=1$}} 
    & \multicolumn{3}{c}{\textbf{$K_\mathrm{est}=2$}} \\
\cmidrule(lr){1-3}
\cmidrule(lr){4-6}
\cmidrule(lr){7-9}
\cmidrule(lr){10-12}
{\scriptsize \textbf{$K_\mathrm{gt}$}} & {\scriptsize \textbf{$N_\mathrm{d}$}} & {\scriptsize {SAR} (dB)}
    & {\scriptsize \textit{MagLS}} & {\scriptsize \textit{COMPASS}} & {\scriptsize \textit{nCOMPASS}}
    & {\scriptsize \textit{MagLS}} & {\scriptsize \textit{COMPASS}} & {\scriptsize \textit{nCOMPASS}}
    & {\scriptsize \textit{MagLS}} & {\scriptsize \textit{COMPASS}} & {\scriptsize \textit{nCOMPASS}} \\
\midrule
0 & 0 & $-\infty$
    & \grey 0.1059 & \grey 0.1266 & \grey \textbf{0.0630}
    & 0.1059 & 0.7965 & \textbf{0.0625}
    & 0.1059 & 1.6676 & \textbf{0.0739} \\
0 & 1 & $-\infty$
    & \grey 0.1461 & \grey 0.3722 & \grey \textbf{0.0630}
    & 0.1461 & 0.9141 & \textbf{0.0641}
    & 0.1461 & 1.9701 & \textbf{0.1104} \\
1 & N/A & $\infty$
    & \textbf{3.8184} & 5.3149 & 6.9306
    & \grey 3.8184 & \grey \textbf{0.0000} & \grey \textbf{0.0000}
    & 3.8184 & \textbf{0.0000} & \textbf{0.0000} \\
1 & 0 & 0
    & \textbf{0.8849} & 1.5234 & 1.9272
    & \grey 0.8849 & \grey 0.3262 & \grey \textbf{0.0621}
    & 0.8849 & 1.0711 & \textbf{0.0598} \\
1 & 1 & 0
    & \textbf{0.8973} & 1.5179 & 1.8368
    & \grey 0.8973 & \grey 0.4221 & \grey \textbf{0.0635}
    & 0.8973 & 1.1066 & \textbf{0.0631} \\
2 & N/A & $\infty$
    & \textbf{2.3879} & 4.0468 & 4.4093
    & \textbf{2.3879} & 2.9840 & 3.8366
    & \grey 2.3879 & \grey \textbf{0.0546} & \grey 0.2204 \\
2 & 0 & 0
    & \textbf{0.7346} & 1.2236 & 1.0815
    & \textbf{0.7346} & 1.0344 & 0.7461
    & \grey 0.7346 & \grey 0.9554 & \grey \textbf{0.0611} \\
2 & 1 & 0
    & \textbf{0.7274} & 1.2177 & 0.9890
    & \textbf{0.7274} & 1.1508 & 0.7731
    & \grey 0.7274 & \grey 0.8945 & \grey \textbf{0.0651} \\
\bottomrule
\end{tabular}
\label{table:ild_rmse} 
\end{table*}

\begin{table*}[t!]
\centering
\caption{IC RMSE values.}
\begin{tabular}{ccc| *{3}{ccc}}
\toprule
\multicolumn{3}{c|}{\textbf{Input sound scene}} 
    & \multicolumn{3}{c}{\textbf{$K_\mathrm{est}=0$}} 
    & \multicolumn{3}{c}{\textbf{$K_\mathrm{est}=1$}} 
    & \multicolumn{3}{c}{\textbf{$K_\mathrm{est}=2$}} \\
\cmidrule(lr){1-3}
\cmidrule(lr){4-6}
\cmidrule(lr){7-9}
\cmidrule(lr){10-12}
{\scriptsize \textbf{$K_\mathrm{gt}$}} & {\scriptsize \textbf{$N_\mathrm{d}$}} & {\scriptsize {SAR} (dB)}
    & {\scriptsize \textit{MagLS}} & {\scriptsize \textit{COMPASS}} & {\scriptsize \textit{nCOMPASS}}
    & {\scriptsize \textit{MagLS}} & {\scriptsize \textit{COMPASS}} & {\scriptsize \textit{nCOMPASS}}
    & {\scriptsize \textit{MagLS}} & {\scriptsize \textit{COMPASS}} & {\scriptsize \textit{nCOMPASS}} \\
\midrule
0 & 0 & $-\infty$
    & \grey 0.2516 & \grey 0.0204 & \grey \textbf{0.0061}
    & 0.2516 & 0.0933 & \textbf{0.0064}
    & 0.2516 & 0.1671 & \textbf{0.0069} \\
0 & 1 & $-\infty$
    & \grey 0.2671 & \grey 0.0219 & \grey \textbf{0.0062}
    & 0.2671 & 0.0978 & \textbf{0.0062}
    & 0.2671 & 0.1723 & \textbf{0.0075} \\
1 & N/A & $\infty$
    & 0.7045 & \textbf{0.5673} & 0.6236
    & \grey 0.7045 & \grey \textbf{0.0000} & \grey \textbf{0.0000}
    & 0.7045 & \textbf{0.0000} & \textbf{0.0000} \\
1 & 0 & 0
    & 0.3357 & \textbf{0.1946} & 0.2194
    & \grey 0.3357 & \grey 0.0497 & \grey \textbf{0.0061}
    & 0.3357 & 0.1175 & \textbf{0.0065} \\
1 & 1 & 0
    & 0.3543 & \textbf{0.1887} & 0.2108
    & \grey 0.3543 & \grey 0.0520 & \grey \textbf{0.0054}
    & 0.3543 & 0.1266 & \textbf{0.0061} \\
2 & N/A & $\infty$
    & 0.4871 & \textbf{0.3701} & 0.4056
    & 0.4871 & \textbf{0.2957} & 0.3693
    & \grey 0.4871 & \grey \textbf{0.0040} & \grey 0.0134 \\
2 & 0 & 0
    & 0.3105 & \textbf{0.1403} & 0.1518
    & 0.3105 & 0.1233 & \textbf{0.1143}
    & \grey 0.3105 & \grey 0.1081 & \grey \textbf{0.0058} \\
2 & 1 & 0
    & 0.3179 & \textbf{0.1386} & 0.1501
    & 0.3179 & 0.1252 & \textbf{0.1097}
    & \grey 0.3179 & \grey 0.1129 & \grey \textbf{0.0061} \\
\bottomrule
\end{tabular}
\label{table:ic_rmse} 
\end{table*}

To gain preliminary insights into the performance of the proposed framework, a variety of sound field scenarios were simulated and captured using both a first-order Ambisonic receiver and a binaural receiver (the latter employing the KU100 HRTFs from \cite{armstrong2018perceptual}). Binaural energy, interaural level difference (ILD), and interaural coherence (IC) were then computed over frequency (see, e.g., Sec.\,VI.A of \cite{fernandez2022enhancing}) for the reference binaural signals, as well as for the binaural signals obtained by rendering the first-order inputs using \textit{nCOMPASS}, \textit{COMPASS} \cite{politis2018compass}, and \textit{MagLS} \cite{schorkhuber2018binaural}. Root-mean-square-error (RMSE) values between the reference and rendered binaural outputs were subsequently calculated and averaged over the Equivalent Rectangular Bandwidth (ERB) scale, yielding a single error value for each tested scenario.

The tested scenarios involved varying numbers of uncorrelated noise sources ($K_\mathrm{gt} \in [0,1,2]$) within a free-field, or an isotropic ($N_\mathrm{d}=0$) or first-order anisotropic ($N_\mathrm{d}=1$) diffuse-field, with different source to ambience ratios (SAR). Since multi-source parametric methods can be sensitive to errors in the estimated number of sources ($K_\mathrm{est}$), the \textit{COMPASS} and \textit{nCOMPASS} renders were conducted when setting $K_\mathrm{est} \in [0,1,2]$. In cases where the estimated number of sources exceeded the true number of sources (i.e., $K_\mathrm{est}>K_\mathrm{gt}$), random DOAs were introduced into the model. Otherwise, exact DOAs were employed for all tests, and the predicted ambience SH order was set to 1st order for all \textit{nCOMPASS} renders (regardless of $N_\mathrm{d}$). 
The above described scenarios were simulated 100 times using randomised DOAs and random anisotropic ambience distributions (when $N_\mathrm{d}>0$), and the final average (over all trails and ERB frequencies) RMSE values are presented in Tables \ref{table:timbre_rmse}-\ref{table:ic_rmse}. Note that in cases where the model matches the input scene (i.e., $K_\mathrm{gt}=K_\mathrm{est}$) the error values are highlighted in blue; otherwise, a mismatch exists between the input scene and adopted model. 

For the highlighted matched conditions, it can be seen that \textit{nCOMPASS} almost always leads to the lowest RMSE values. Additionally, \textit{nCOMPASS} generally fares better than \textit{COMPASS} when the input scene contains both sources and ambience ($N_\mathrm{d}\geq 0$); perhaps owing to its more flexible sound field model, avoiding intermediate signal estimates and beamforming, and its optimal rendering with cue matching. When there is an underestimation of the number of sources ($K_\mathrm{est}<K_\mathrm{gt}$), regular \textit{COMPASS} generally leads to the lowest IC errors and \textit{MagLS} leads to the lowest timbral colouration and ILD errors. Whereas, when there is an overestimation of the number of sources ($K_\mathrm{est}>K_\mathrm{gt}$), \textit{nCOMPASS} produces notably lower error values than both \textit{COMPASS} and \textit{MagLS} is the majority of cases. Therefore, adopting source number estimators that err on the side of overestimating $K_\mathrm{est}$, may lead to better performance. However, to gain a more robust understanding of the perceptual performance of the proposed rendering framework, formal listening tests involving realistic sound scenes are required.

\section{Perceptual evaluation}
\label{sec:evaluation}

A binaural multiple-stimulus listening test was conducted, which involved comparing ground-truth simulated binaural sound scenes against renderings produced by the proposed framework, and also by closely related parametric methods \cite{berge2010high,politis2015sector,mccormack2022estimating,schorkhuber2019linearly} and the MagLS approach \cite{schorkhuber2018binaural,deppisch2021end}.

\subsection{Implementation}
\label{sec:implementation}

The framework proposed in this study was implemented and realised as an audio plugin\footnote{\url{https://github.com/leomccormack/KOMPASSI-Renderer-Plugin}}, developed using JUCE\footnote{\url{https://github.com/juce-framework/JUCE}} and the Spatial\_Audio\_Framework\footnote{\url{https://github.com/leomccormack/Spatial_Audio_Framework}}. The selected time-frequency transform was the alias-free STFT \cite{vilkamo2017time}, which has been used by the present authors for developing other related parametric methods in previous studies \cite{politis2017enhancement,mccormack2022estimating}. The implementation of the framework was divided into separate encoder and decoder blocks (see Fig.~\ref{fig:block_diagram}). The encoder accepts either Ambisonic sound scenes or raw microphone array recordings as input. The SCMs are computed over time windows of length 11\,ms (512 samples at 48\,kHz), with the parameter estimators used in \cite{politis2018compass,mccormack2022parametric} adopted for the present study; namely: SORTE \cite{han2013improved} for the source number detection, and MUSIC \cite{schmidt1986multiple} for the DoA estimation. Note that SORTE always produces $K_\mathrm{est}\geq 1$, and has a general tendency to overestimate the number of sources \cite{mccormack2023spatial}. The individual source powers and the SHD ambience power distribution are estimated as described in Sec.~\ref{sec:estModelParams}. The ambience estimation order was set to $\mathrm{min}( \lfloor \sqrt{M}-1 \rfloor,2)$.
The decoder takes the input signals and accompanying spatial parameters as input, and uses them to generate output signals corresponding to the target setup using the SCM matching solution described in Sec.~\ref{sec:renderingMethods}. These mixing matrices are updated every 11\,ms, and are recursively averaged using a one-pole filter with a coefficient value $0.8$. Decorrelation was performed using a combination of fixed frequency-dependent delays and cascaded lattice all-pass filters, with longer delays and higher filter orders applied to the lower sub-bands.

\subsection{Simulation of the test sound scenes}

The test sound scenes were simulated\footnote{\url{https://github.com/polarch/shoebox-roomsim}} using the image‑source method \cite{allen1979image}. Four scenarios were created. The first, \textbf{anechoic\_mix}, comprised four sources (clapping, water fountain, piano, female speech) positioned on the horizontal plane at azimuths $[90, 30, -30, -90]$ in a free field. The second, \textbf{small\_speech}, contained two male and two female speakers at the same azimuths inside a small shoebox room ($6 \times 5 \times 3.1$\,m), with all sources one metre from the listener located near the room centre. The RT60 values were $[0.33, 0.39, 0.26, 0.20, 0.07, 0.04]$\,s across octave bands from $125$\,Hz to $4$\,kHz. The third scenario, \textbf{hall\_mozart}, recreated an orchestral performance using anechoic instrument recordings from \cite{patynen2008anechoic}. The ensemble (six violins, three violas, three cellos, one double bass, bassoon, timpani, two flutes, two clarinets, and a five‑instrument brass section) was placed in a large shoebox room ($35.7 \times 19.8 \times 17.4$\,m) emulating the Musikvereinssaal in Vienna. RT60 values were $[2.77, 2.93, 2.91, 2.93, 2.66, 1.90, 1.15]$\,s from $125$\,Hz to $8$\,kHz, and instrument positions followed a standard orchestral layout. The fourth scenario, \textbf{medium\_band}, featured bass guitar, drums, shaker, and synthesised strings at azimuths $[90, 30, -30, -90]$ inside a medium‑sized room ($10 \times 6 \times 3.1$\,m). All sources were one metre from the listener, with RT60 values of $[0.52, 0.59, 0.39, 0.20, 0.16, 0.13]$\,s across octave bands from $125$\,Hz to $4$\,kHz.

Ground‑truth binaural reference signals were generated by convolving the direction‑dependent source signals with the nearest HRTFs from an 8802‑point KU100 dataset \cite{armstrong2018perceptual}, using metadata extracted from the simulated echograms. This approach enables substituting HRTFs with SH vectors or ATFs, allowing Ambisonic or microphone‑array recordings of the same scenes to be simulated and subsequently rendered using the methods under test.

\subsection{Listening test design}

\begin{figure*}[!t]
\centerline{
\includegraphics[trim=0cm 0cm 0cm 0,clip=true,height=8.7cm]{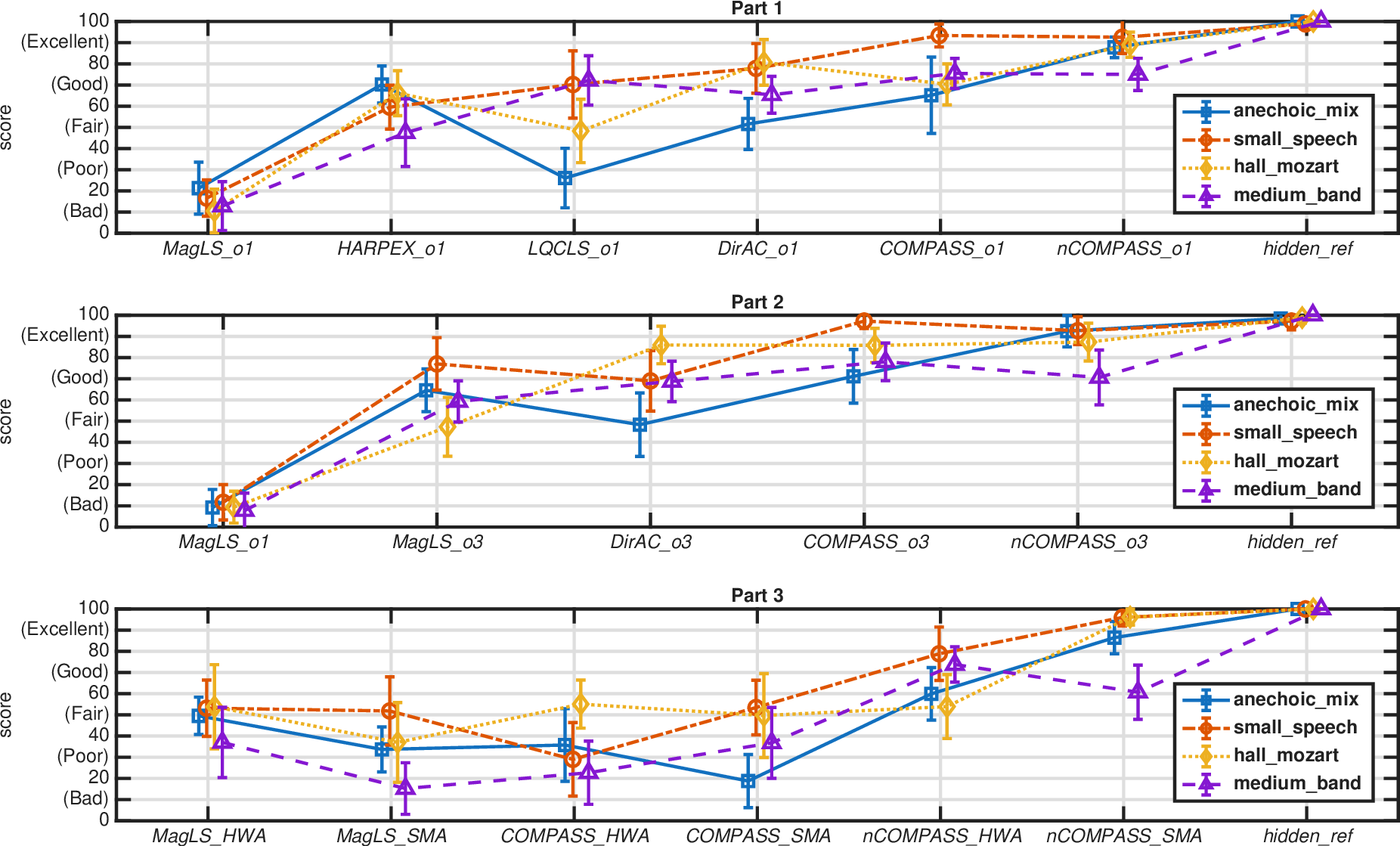}
}
\caption{Medians and 95\% confidence intervals for listening test parts 1, 2 and 3. }
\label{fig:listeningTestResults}
\end{figure*}

The listening test comprised three parts, with each part comparing the proposed framework against state-of-the-art parametric methods tailored to a specific input format. Only published methods with mature, publicly available real-time implementations provided by their original authors were included as comparison baselines. The first part evaluated nCOMPASS using FOA input, which remains the most widely adopted spatial audio capture format due to the broad availability of tetrahedral Ambisonic microphones. FOA versions of the test scenes were synthesised and reproduced binaurally using the following decoders: the proposed method (\textit{nCOMPASS\_o1}); COMPASS \cite{politis2018compass} (\textit{COMPASS\_o1}) using the covariance-matching ambience reproduction of \cite{mccormack2022estimating} as implemented in SPARTA (v1.6.2) \cite{mccormack2019sparta}; DirAC (\textit{DirAC\_o1}) using the optimal-mixing formulation of \cite{politis2017enhancement} as also implemented in SPARTA (v1.6.2); the IEM AdaptiveBinauralDecoder\footnote{\url{https://plugins.iem.at/docs/adaptivebinauraldecoder/}} (\textit{LQCLS\_o1}) \cite{schorkhuber2019linearly,giller2019super}; and HARPEX \cite{berge2010high} (\textit{HARPEX\_o1}) via the HARPEX-X plugin\footnote{\url{https://harpex.net/}} (v1.6). MagLS (\textit{MagLS\_o1}) \cite{schorkhuber2018binaural}, as also implemented in SPARTA (v1.6.2), served as a signal-independent baseline. All FOA decoders feature SOFA file readers \cite{majdak2022spatially}, which facilitated loading in the same KU100 HRTFs \cite{armstrong2018perceptual} as used for generating the binaural references.

The second part evaluated the proposed method and existing parametric approaches using HOA (third-order) input. The HOA scenes were rendered using \textit{nCOMPASS\_o3}, \textit{COMPASS\_o3}, and \textit{DirAC\_o3}. First- and third-order MagLS (\textit{MagLS\_o1}, \textit{MagLS\_o3}) served as an anchor and signal-independent baseline condition, respectively. All cases were rendered using the same publicly available implementations as used for the FOA renders, and used the same KU100 HRTFs.

The third part assessed performance when operating directly in the space domain using head-worn microphone array (HWA) recordings, motivated by recent interest in AR audio capture and reproduction \cite{lubeck2022perceptual,mccormack2022parametric,stahl2024perceptual}. Array recordings were generated using the 6-channel HWA ATFs from the EasyCom dataset \cite{donley2021easycom}. These were rendered binaurally using the proposed method (\textit{nCOMPASS\_HWA}); a space-domain variant of COMPASS (\textit{COMPASS\_HWA}) adapted from \cite{mccormack2022parametric} to directly target binaural output; and a space-domain MagLS formulation \cite{deppisch2021end,lubeck2022perceptual} (\textit{MagLS\_HWA}). Implementations of all three methods are available in the accompanying audio plugin$^1$. For additional insight, the same three methods were also applied to a simulated 6-channel SMA (rigid baffle radius 0.2\,cm, uniform sensor placement), yielding \textit{nCOMPASS\_SMA}, \textit{COMPASS\_SMA}, and \textit{MagLS\_SMA}. All renderings again used the same KU100 HRTFs.

A hidden reference (\textit{hidden-ref}) was included in all test parts. Listening tests were conducted in dedicated booths located at Aalto University, Finland, with a background noise level of $\mathrm{L_{A,eq,30\,s}} = 22.0$\,dB SPL(A), using Sennheiser HD600 headphones. The rating scale ranged from 0 to 100 with verbal anchors (bad, poor, fair, good, excellent) at 20-point intervals. Subjects could freely switch between test cases (presented in randomised order), loop shorter segments, and listen for as long as needed before making their assessments. The average time taken to complete all three parts of the test was approximately 45\,minutes.  Thirteen participants took part. All participants reported normal hearing, had prior experience in perceptual testing, and had no knowledge of the study hypothesis.

\subsection{Listening test results and statistical analysis}
\label{sec:results}

The listening test results for all three parts of study are presented in Fig. \ref{fig:listeningTestResults}. A subset of ratings (reference, nCOMPASS, COMPASS, and MagLS) was analysed to determine whether differences were statistically significant. Normality was rejected using the Jarque--Bera test, so a non-parametric Friedman test with Wilcoxon signed-rank post-hoc comparisons was applied. Bonferroni correction yielded an adjusted significance level of $\alpha = 0.0167$.

\subsubsection{Part 1 (o1 conditions)} Significant differences were found for the \textbf{anechoic\_mix} scene ($\chi^2(2,13)=17.91, p<0.001$) and \textbf{hall\_mozart} ($\chi^2(2,13)=20.17, p\ll0.001$), with all nCOMPASS pairwise comparisons significant in both scenes (e.g., anechoic\_mix: $W=0, p=0.002$ vs.\ reference; $W=70, p=0.011$ vs.\ COMPASS; $W=91, p<0.001$ vs.\ MagLS). For \textbf{medium\_band}, the Friedman test was significant ($\chi^2(2,13)=18.63, p\ll0.001$), with nCOMPASS differing significantly from the reference and baseline, but not from COMPASS. No significant differences were found for \textbf{small\_speech}.

\subsubsection{Part 2 (o3 conditions)} The same analysis was applied to nCOMPASS\_o3, COMPASS\_o3, MagLS\_o3, and the reference. Significant differences were found for \textbf{anechoic\_mix} ($\chi^2(2,13)=17.64, p<0.001$), \textbf{medium\_band} ($\chi^2(2,13)=10.65, p=0.005$), and \textbf{hall\_mozart} ($\chi^2(2,13)=18.96, p\ll0.001$). Across these scenes, nCOMPASS differed significantly from COMPASS and the baseline, and (except for medium\_band) also from the reference. No significant differences were found for \textbf{small\_speech}.

\subsubsection{Part 3 (HWA and SMA conditions)} With six pairwise comparisons, the adjusted significance level was $\alpha = 0.0083$. Friedman tests were significant for all scenes: \textbf{anechoic\_mix} ($\chi^2(2,13)=24.78, p<0.001$), \textbf{small\_speech} ($\chi^2(2,13)=26, p\ll0.001$), \textbf{hall\_mozart} ($\chi^2(2,13)=19.54, p\ll0.001$), and \textbf{medium\_band} ($\chi^2(2,13)=24.15, p\ll0.001$). Post-hoc tests showed five significant comparisons in \textbf{anechoic\_mix}, all six in \textbf{small\_speech}, five in \textbf{medium\_band}, and three in \textbf{hall\_mozart}.

\subsection{Results discussion}
\label{sec:discussion}

Generally, sound scenes containing only a few temporally and spectrally overlapping sources in dry or anechoic conditions are known to challenge parametric methods \cite{politis2017enhancement}. Such scenes often violate the assumed sound field models, leading to audible artefacts, particularly when spatial parameter estimation becomes unreliable at low Ambisonic orders or with arrays comprising few microphones. Low reverberation levels further reduce perceptual masking, making these artefacts more noticeable. Although temporal averaging is commonly used to stabilise parameter estimates, selecting a window that suits both transient and stationary content is inherently difficult. Within this context, the \textbf{anechoic\_mix} scene represents one of the most demanding scenarios. Despite this, nCOMPASS consistently received ratings near the upper end of the evaluation scale across all test parts. Statistical analysis confirmed significant improvements over COMPASS in every comparison, and nCOMPASS also outperformed MagLS in nearly all cases, with the only non-significant result occurring for the head-worn array.

The \textbf{small\_speech} scene, although containing four simultaneous talkers, is less problematic due to the temporal and spectral sparsity of speech. In Parts~1 and~2, nCOMPASS and COMPASS produced similar ratings, and no significant differences were detected. Both parametric methods clearly outperformed the MagLS baselines. In Part~3, however, nCOMPASS was rated significantly higher than both COMPASS and MagLS for the corresponding arrays. For the \textbf{hall\_mozart} scene, where the low direct-to-reverberant ratio emphasises the rendering of diffuse ambience, nCOMPASS again produced some of the highest ratings across all test parts. Pairwise comparisons showed significant improvements in most cases. The \textbf{medium\_band} scene, which includes broadband transient sources such as shaker and drum kit, also presents a challenging scenario. When using Ambisonic receivers, nCOMPASS did not yield a noticeable improvement over COMPASS. However, when operating directly in the space domain using either the HWA or SMA, nCOMPASS achieved statistically significant improvements over both COMPASS and MagLS.

\section{Conclusion}
\label{sec:conclusion}


This article presented nCOMPASS, a unified framework for the parametric analysis of spatial sound scenes captured with arbitrary recording setups and reproduced over arbitrary playback setups. It achieves this through a symmetric representation of capture and playback signals and their inter-channel statistics as realizations of the same sound scene components, spatially encoded by capture- and playback-related transfer functions, respectively. The representation also naturally compensates for independent, known capture and playback setup rotations. The method extends earlier parametric approaches by adopting a flexible time-frequency scene representation comprising a variable number of primary source components, each described by direction-of-arrival and power parameters, together with an anisotropic ambience component represented by a continuous angular power distribution described by spherical harmonic coefficients. A corresponding parameter-estimation procedure was developed to fit the model to the observed spatial covariance structure of the recorded sound field. The method is robust to closely spaced primary source components, while the ambient-field representation captures residual spatial structure that cannot be adequately described by a small number of discrete sources. The model parameters are used to model the playback signal inter-channel statistics which include all relevant spatial cues for reproduction. An optimal mixing matrix that directly transcodes the capture signals to the playback signals is derived. The solution meets the target spatial cues and avoids intermediate spatial filters that can cause artifacts in challenging scene scenarios. Subjective listening tests demonstrated that nCOMPASS achieves improved perceptual accuracy over state‑of‑the‑art linear baselines and existing parametric methods across a range of receiver geometries and sound‑scene types.

\bibliographystyle{IEEEtran}
\bibliography{refs.bib} 

\newpage

 




\vfill

\end{document}